\documentstyle[aps,epsfig]{revtex}
\begin{document}

\draft

\title{Functional Schr\"odinger Picture for
Conformally Flat Space-Time with Cosmological Constant
\thanks{
The talk given at the V International Conference on Gravitation
and Astrophysics of Asian - Pacific Countries, October 1-7 2001, Moscow,
Russia.
}}

\author{
Yuri G. Palii~\thanks{
E-mail: palii@thsun1.jinr.ru}~\thanks{
Permanent address:
Institute of Applied Physics, Moldova Academy of Sciences,
Chisinau, Republic of Moldova, MD-2028.
}\\
\small\it { Laboratory of Information Technologies,}\\
\small\it {Joint Institute for Nuclear Research,}
\small\it {141980, Dubna, Moscow Region, Russia}}

\date{\today}

\maketitle

\begin{abstract}
A quantum-field model of the conformally flat space is formulated
using a standard field-theoretical technique, a probability interpretation and
a way to establish the classical limit. The starting point is the following:
after conformal transformation of the Einstein -- Hilbert action, the
conformal factor represents a scalar field with the negative kinetical term
and the self-interaction inspired by the cosmological constant.  (It has been
found that quanta of such action have a negative value as a sequence of the
negative energy.) The metric energy-momentum tensor of this scalar field is
proportional to the Einstein tensor for the initial metric.  Therefore, a
vacuum state of the field is treated as a classical space. In such vacuum the
zero mode is a scale factor of the flat Friedmann Universe. It is shown that
conformal factor may be viewed as a an inflaton field, and its small
non-homogeneities represent gauge invariant scalar metric perturbations.
\end{abstract}



\section{Introduction}


The metric of the conformally flat space-time only contains one essential
variable -- a conformal factor~\cite{Sing} which is a scale factor of the
expanding Universe. The negative energy of this variable leads to the
difficulties in the quantum cosmology based on the Wheeler -- DeWitt equation
\footnote
{The reduced phase space quantization~\cite{Isham} allows one to ignore the
negative sign if the scale factor remains a single degree of freedom.
A special variable must be presented in extended phase space
to become a time-like parameter after reduction. If such a variable was the
scale factor, a possibility of a study on the quantum properties of the early
Universe (for example,~\cite{Narlik1,Pad1}) would disappear.}
~\cite{DeWitt} (for example, the wave function of the Universe is
non-normalazible, a probability current density is not positive definite).
Different ways are used to overcome these obstacles. Matter or scale factor
sometimes are treated as non-dynamical. Neglecting an interaction
between matter and scale factor allows one to factorize the wave function and
to include a minus sign into the time or the energy in the stationary
Schr\"odinger equation for scale factor.  For example, Narlikar and
Padmanabhan~\cite{NarPad} introduce stationary states of the quantum geometry
as a solution of such an equation to prevent classical singularity.
However, in this approach the conformal factor is treated as a physical
variable while its energy is considered as an unphysical one. Hence it seems
rather strange that this energy may compensate the energy of the created
matter~\cite{Pad1}.  Another interesting result of the quantization of the
conformal factor via Feynman path integral derived by these authors is
the growing of the quantum fluctuations near the singularity~\cite{NarPad}.
But the Euclidian path integral for gravity is divergent because the Euclidian
gravitational action is not positive-definite~\cite{Hawking}.  Gravitational
waves carry positive energy and gravitational potential energy is negative.
Conformal rotation (using the complex conformal factor~\cite{Gibbons}) makes
an Euclidian path integral convergent by changing the sign of the action
\footnote{
A related problem is that the sign of the Euclidian action for
scale factor must be changed with respect to the matter case to get a
physically accepted probability for the birth of the Universe~\cite{LindeJ}.}.
But this method is applicable only if the conformal factor is an unphysical
variable (as in asymptotically flat space-time)~\cite{Schleich}
\footnote{
This condition also constrains validity of a conclusion about
positivity of the gravitational energy~\cite{Weinbergbook}. Moreover, the
mentioned conclusion concerns the energy of gravitational waves which do not
describe the expansion of the Universe.}.

In Friedmann -- Robertson -- Walker Universe, the conformal factor is a single
degree of freedom~\cite{DeWitt}. It must be a physical variable
because the expansion of the Universe is an observational fact
\footnote{Theoretical argumentation why the conformal factor is a physical
degree of freedom is done in~\cite{Ryan,Pad15,Pad19}.}.

To conclude the brief review of theoretical works dealing with quantization
of the conformal factor, one can say that there is no completely satisfying
quantum picture of the expansion of the Universe.
The main questions are gravitational degrees of freedom and gravitational
energy.

Experimental data point out an accelerating expansion of the Universe,
and hence, the cosmological constant may carry a significant contribution
into the energy density~\cite{Perl}. It is a difficult question to the quantum
field theory why the expected vacuum energy density is much greater than that
corresponding to the cosmological constant~\cite{WeinbL}. Inflation theories
exploit a state equation related to the cosmological constant but they do not
explain the mentioned contradiction~\cite{Kolb,Gonzalez}. Meanwhile, the large
scale structure of the Universe and the cosmic background radiation anisotropy
are treated as consequences of quantum fluctuation of inflaton
field~\cite{Lyth} or quantum metric fluctuation~\cite{Halliwell}
\footnote{
Scalar metric perturbations play an important role in theoretical explanation
of the CMB anisotropy due to Sachs -- Wolfe effect~\cite{Sachs}. As in quantum
field theory matter is described by 4-covariant quantities, quantum
fluctuations of the conformal factor (4-scalar) may be interesting as a kind
of metric perturbations.  }.
A modern tool for studying the inflation is a
non-equilibrium phase transition dynamics~\cite{Boyan} in the field
theoretical Schr\"odinger picture~\cite{Jackiw}.

So the quantum model of the conformally flat space-time looks interesting
from a theoretical viewpoint and may link quantum cosmology to the
observational data.

In this paper, attempt is made to quantize the conformally flat
space-time in the Schr\"odinger picture by a standard field theoretical
technique.  Two main circumstances are taken into account. First,
the geometrical and field properties of the space-time are split in the
Einstein -- Hilbert action via a conformal transformation. The equation of
motion for the gravitational field (conformal factor) is derived by variation
of the action with respect to a field variable and represents a Klein --
Gordon equation with additional Penrous -- Chernicov -- Tagirov
term~\cite{Chern}.  The metric energy-momentum tensor~\cite{Chern} of the
field is derived by variation of the action with respect to new metric
(Minkowcki metric for conformally flat space-time discussed here)
and is proportional to Einstein equations for the initial metric.
Thus a vacuum state of the field
(where the energy-momentum tensor overage is zero)
is considered as a classical space-time.
Second, it is taken into account that a system with negative energy needs to
be quantized using Planck constant with a negative sign according to the
Bohr -- Sommerfeld rule and to analogy between wave mechanic and geometric
optics.

The paper is organized as follows. Section~\ref{CF} represents
the conformally flat space-time with the cosmological constant as a
self-interacting scalar field in Minkowski space-time.
In Section~\ref{SP} quantization of the oscillating modes is fulfilled
while zero mode is considered as a classical quantity.
Section~\ref{Rel} gives a relation of the conformal factor to an inflaton field
and to scalar metric perturbations. The Summary reviews the
construction of the model, discusses main results,
and compares with some related works. Quantization of the system
with the negative energy (inverted harmonic oscillator) is prepared
in Appendix~\ref{ApA}. Some used formulae for Bessel functions are given
in Appendix~\ref{ApB}.


\section{Representation of the conformally flat space-time by a scalar field}
\label{CF}


To describe the conformally flat Universe by a scalar field in Minkowski
space-time, we use conformally flat coordinates. Let us suppose that there is
a general coordinate transformation

\begin{equation} \label{gct}
x=x(\tilde x),\;\; 
g_{ij}(x)=\frac{\partial \tilde x^l}{\partial x^i}
\frac{\partial \tilde x^m}{\partial x^j} \tilde g_{lm}(\tilde x),
\;\;\;\;i,j=1,2,3,4
\end{equation}

\begin{equation} \label{dscf}
ds^2=g_{ij}(x)dx^idx^j
=\tilde g_{ij}(\tilde x)d{\tilde x}^id{\tilde x}^j,
\end{equation}
after which the new metric has a common multiplier (conformal factor):

\begin{equation} \label{metric}
\tilde g_{ij}(\tilde x)=a^2(\tilde x){g'}_{ij}(\tilde x),
\end{equation}

\begin{equation} \label{interval}
ds^2=a^2(\tilde x)(ds')^2,\;\;\;\;
(ds')^2={g'}_{ij}(\tilde x)d\tilde x^id\tilde x^j.
\end{equation}
The equations for metric $\tilde g_{ij}(\tilde x)$ following from
Einstein -- Hilbert action
\footnote{
Notations correspond to Landau, Lifshitz book
"Classical Theory of Fields"~\cite{Landau}. Geometric units
system~\cite{Misner} $16\pi G=c=1$ ($G$ is the Newtonian constant and $c$ is
speed of light) is used.  Conformal factor $a(x)$ has dimension of the
length, and cosmological constant $\Lambda$ has a dimension of the length in to
the minus square. It is convenient
to have the Planck constant in formulae apparently when quantization
is discussed. In usual for quantum gravity system of units $16\pi G=c=\hbar=1$
all the quantities, whose dimension expresses in terms of mass, length and time
units, become dimensionless.}

\begin{equation} \label{EH}
S[\tilde g(\tilde x)]=-\int\limits_{}^{}d^4\tilde x\sqrt{-\tilde g}\left(\tilde R+\Lambda\right),\;\;\;\;
\Lambda>0,
\end{equation}
can be written as equations for conformal factor $a(\tilde x)$ and metric
${g'}_{ij}(\tilde x)$
\footnote{
If one considers the conformal factor as a new variable, an additional
condition must be added (for example, fixing the scalar curvature or
determinant of the metric) to preserve the number of independent variables.}

\begin{eqnarray}
\tilde G_{ij}&=&\tilde R_{ij}(\tilde x)
-\frac{1}{2}\tilde R(\tilde x)\tilde g_{ij}(\tilde x)
-\Lambda\tilde g_{ij}(\tilde x)\label{G}\\
&=&{R'}_{ij}(\tilde x)-\frac{1}{2}R'(\tilde x){g'}_{ij}(\tilde x)\nonumber\\
&&+a^{-2}(\tilde x)[-\nabla_i\nabla_j a^2(\tilde x)
+6\partial_ia(\tilde x)\partial_ja(\tilde x)\nonumber\\
&&+{g'}_{ij}(\tilde x)(\Box a^2(\tilde x)
-3\partial^na(\tilde x)\partial_na(\tilde x)-\Lambda a^4(\tilde x))].
\nonumber
\end{eqnarray}
From the other hand, the Einstein -- Hilbert action~(\ref{EH}) can be rewritten
by using the conformal transformation formulae~\cite{Eizen}
corresponding to factorization of the metric~(\ref{metric})

\begin{equation}
\tilde R(\tilde x)=a^{-2}\left(R'(\tilde x)
-6a^{-1}(\tilde x)\Box a(\tilde x)\right),\;\;\;\mbox{где}\;\;\;
\Box a(\tilde x)=\frac{1}{\sqrt{-g'}}
\partial_i\left(\sqrt{-g'}(g')^{ij}\partial_j a\right)
\end{equation}

\begin{equation}
\sqrt{-\tilde g(\tilde x)}=a^4(\tilde x)\sqrt{-g'(\tilde x)}.
\end{equation}
As a result, an action for a scalar field $a(\tilde x)$ with negative kinetic
term in metric ${g'}_{ij}(\tilde x)$~\cite{Hawking} appears

\begin{equation} \label{csfact}
S[g'(\tilde x),a(\tilde x)]
=-\int\limits_{}^{}d^4{\tilde x}
\sqrt{-g'}
\left(\frac{1}{2} (g')^{ij}\partial_i a\partial_j a
+\frac{1}{12}a^2R'+\frac{\lambda}{4!}a^4 \right),
\end{equation}
where $\lambda=4\Lambda$. Total divergence and numerical factor are
neglected. Variation of this action with respect to metric ${g'}_{ij}$
gives a metric energy-momentum tensor~\cite{Chern} of the conformal scalar
field $a(\tilde x)$

\begin{equation} \label{T}
T^{(metr)}_{ij}=T^{(can)}_{ij}-\frac{1}{6}\left({R'}_{ij}+{g'}_{ij}\Box
-\nabla_i\nabla_j\right)a^2,
\end{equation}
where the canonical energy-momentum tensor has the form

\begin{equation}
T^{(can)}_{ij}=-\partial_i a\partial_j a
+{g'}_{ij}\left(\frac{1}{2}\partial^n a\partial_n a
+\frac{1}{12}a^2R'+\frac{\lambda}{4!}a^4\right).
\end{equation}
The metric energy-momentum tensor~(\ref{T}) is proportional to the
Einstein equations~(\ref{G})

\begin{equation} \label{TG}
T^{(metr)}_{ij}=-\frac{a^2}{6}\tilde G_{ij}.
\end{equation}
This fact can be directly deduced if we mention that both quantities
are derived by variation of the Einstein -- Hilbert action
with respect to the metrics which are proportional each other as it follows
from~(\ref{metric})
\footnote
{The author is indebted to Professor N.A. Chernikov for this remark.}.
The metric energy-momentum tensor
trace is equal to zero on the equation of motion for $a(\tilde x)$

\begin{equation} \label{sled}
\frac{1}{6}a^3\tilde G=-a^{-1}T=\Box a-\frac{1}{6}aR'-\frac{\lambda}{6}a^3.
\end{equation}
So this equation of motion expresses a condition for conformal invariance.
The interaction that arises due to the cosmological constant is a unique
possible conformally invariant one.
We see from~(\ref{TG}) that the metric energy-momentum tensor~(\ref{T})
must be equal to zero to be consistent with Einstein equations~(\ref{G}).
Let us suppose that there is a state in quantum field theory of the
conformal factor $a(x)$ with a vanishing averaged value for the energy-momentum
tensor

\begin{equation} \label{Tvac}
\langle 0|T_{ij}|0\rangle =0, \;\;\;\langle 0|0\rangle=1.
\end{equation}
Such a state naturally is a vacuum state and corresponds to the classical
theory.

In the case of the conformally flat space-time we can consider
the transformation~(\ref{gct})
as a transformation to the conformally flat coordinates $\tilde x$
\footnote
{The explicit form of such transformation for Robertson -- Walker
metric is done in the textbook by Lightman at al.~\cite{Lightman},
problem 19.8.}.
Hence, the metric ${g'}_{ij}(\tilde x)$ arising due to factorization of the
${\tilde g}_{ij}(\tilde x)$~(\ref{metric}) is Minkowski one:

\begin{equation}  \label{Mink}
{g'}_{ij}(\tilde x)=\eta_{ij}=(+,-,-,-),
\end{equation}
and the curvature depending terms disappear from the action~(\ref{csfact}) and
from the energy-momentum tensor~(\ref{T})

\begin{equation} \label{actmin}
S[a(x)]=-\int\limits_{}^{}d^4x\left(\frac{1}{2}\partial^n a\partial_n a
+\frac{\lambda}{4!}a^4\right),
\end{equation}

\begin{equation} \label{tflat}
T_{ij}=-\partial_i a\partial_j a
+\frac{1}{6}\partial_i \partial_j a^2
+\eta_{ij}\left(-\frac{1}{6}\Box a^2
+\frac{1}{2}\partial^n a\partial_n a
+\frac{\lambda}{4!}a^4\right).
\end{equation}
The energy-momentum tensor~(\ref{tflat}) differs from the canonical one by
second derivative terms. But the corresponding Hamiltonian may be derived
in usual form after removing a total divergence.

The solution to the Einstein equations~(\ref{G}) is the de Sitter
space-time with the curvature scalar $R=-\lambda$. In coordinates $\tilde x$
it looks like the spatially flat Friedmann -- Robertson -- Walker Universe
with a cosmological constant.
Isometries of the conformally flat space-time
preserve the factorization of the metric
$\tilde g_{ij}(\tilde x)$ and
action~(\ref{csfact})\footnote
{The conformal Killing vectors of the Robertson -- Walker metric
in conformally flat coordinates are presented in~\cite{Keane}.}.
All conformally flat
spaces have the same conformal group symmetry
~\cite{Dubr}. In particular,
de Sitter group generators undergo to Poincar\'e group ones in the limit
of the large curvature radius~\cite{Gurs} (small cosmological constant).
A correct description of the classical Universe will be guaranteed by a
conserved energy-momentum tensor providing that the field $a(x)$ has a non-zero
vacuum averaged value $a(\eta)$ (it will be shown that $\eta$ corresponds to
conformal time in the classical limit)
playing the role of a scale factor for the classical metric

\begin{equation}
a(\eta,x)=a(\eta)+\delta a(\eta,x),
\end{equation}
where
\begin{equation} \label{aggda}
a(\eta)\gg|\delta a(\eta,x)|.
\end{equation}
Vacuum fluctuations will cause increase of $a(\eta)$ due to instability of
the system with the action~(\ref{actmin}).

The result of the analysis is the following. After factorization of
the conformal factor, the Einstein -- Hilbert action looks like a conformal
scalar field action with the negative kinetic term. The Einstein equations for
the initial metric ~(\ref{G}) are proportional to the metric energy-momentum
tensor of the field~(\ref{T}). Hence, the classical conformally flat
space-time can be represented as a vacuum state of the unstable scalar field
in Minkowski space-time. The energy-momentum tensor vacuum average must be
zero and the vacuum field average is a classical scale factor.


\section{Field-theoretical Schr\"odinger picture}
\label{SP}


The most suitable approach to quantization of the
model~(\ref{actmin}),~(\ref{tflat}) is the Schr\"odinger picture which has
advantages in time-dependent problems~\cite{Jackiw}.
The aim of the quatization is to find a ground state of the system and
to construct creation and annihilation operators. This is done in the way
proposed Guth and Pi~\cite{GuthPi} where the inflaton field with
self-interaction $\lambda\phi^4$ was quantized (in the de Sitter metric)
to study the ``slow-rollover" phase transition in the new
inflation model.
Here the problem has been formulated in Minkowski metric but due to
instability of the system, the field zero mode dynamics leads to an effect
appearing in the non-stationary metric. The field $a(x)$ is decomposed into
the Fourier series

\begin{equation} \label{decompos}
a(\eta,x)=\sqrt{2}a_o(\eta)
+\sum_{\vec k}^{}[a^+_{\vec k}(\eta)\cos\vec k\cdot\vec x
+a^-_{\vec k}(\eta)\sin\vec k\cdot\vec x],
\end{equation}
in a cub with a side $b=2\pi$ using real functions with the following
properties

\begin{equation}
a^+_{-\vec k}(\eta)\equiv a^+_{\vec k}(\eta),\;\;\;\;
a^-_{-\vec k}(\eta)\equiv -a^-_{\vec k}(\eta).
\end{equation}
The approximation

\begin{equation}  \label{aoggak}
a_o(\eta)\gg a^\pm_{\vec k}(\eta)
\end{equation}
corresponds to the classical limit condition~(\ref{aggda}). Further
simplification is to neglect the interaction between the modes
$a^\pm_{\vec k}(\eta)$ in virtue of a small value of $\lambda$. Then the
action~(\ref{actmin}) takes the form

\begin{equation} \label{S}
S=\int\limits_{}^{}d\eta\left\{-\dot
a^2_o(\eta)-\frac{\lambda}{3!}a^4_o(\eta)
+\sum_{k}^{}\left[-\dot a^2_k(\eta)+(k^2-\lambda
a^2_o(\eta))a^2_k(\eta)\right]\right\},
\end{equation}
where only one sign modes are presented, their notation is simplified
and a space integration is prepared.

\subsection{\it Solution of a classical problem}

The action~(\ref{S}) leads to equations of motions

\begin{equation} \label{scc}
\left\{\begin{array}{l}
\ddot a_o(\eta)-\frac{\lambda}{3}a^3_o(\eta)
-\lambda a_o(\eta)\sum_{k}^{}a^2_k(\eta)=0,\\
\\
\ddot a_k(\eta)+(k^2-\lambda a_o^2(\eta))a_k(\eta)=0,
\end{array}\right.
\end{equation}
which self-consistently describe independent oscillators
$a_k(\eta)$ in external field $a_o(\eta)$. Not self-consistent analitically
solvable problem does not take into account the $a_k(\eta)$-modes influence
on zero mode

\begin{equation} \label{aoak}
\left\{\begin{array}{l}
\ddot a_o(\eta)-\frac{\lambda}{3}a^3_o(\eta)=0,\\
\\
\ddot a_k(\eta)+(k^2-\lambda a_o^2(\eta))a_k(\eta)=0.
\end{array}\right.
\end{equation}
The first integral for a zero mode equation

\begin{equation}
\dot a_o^2(\eta)-\frac{\lambda}{6}a^4_o(\eta)=0.
\end{equation}
must be equal to zero according to the correspondence condition~(\ref{TG}).
The second integration constant can be fixed by assumption that the origin
of the time $\eta$ (the conformal time for the
Friedmann -- Robertson -- Walker interval~\cite{Misner}) coincides with the
origin of the proper time in synchronous coordinate system

\begin{equation} \label{tpr}
dt_{pr}=\sqrt{2}a_o(\eta)d\eta.
\end{equation}
So $a_o(\eta)$ is a scale factor of the flat
Friedmann Universe with the cosmological constant

\begin{equation}   \label{aot}
a_o=\frac{\sqrt{6/\lambda}}{1-\eta}
=\sqrt{\frac{6}{\lambda}}\exp\left\{\sqrt{\frac{\lambda}{12}}t_{pr}
\right\}
\end{equation}
considered in the intervals

\begin{equation}
\begin{array}{rcl}
0\leq & \eta & <1\\
0\leq & t_{pr} & <\infty\\
\sqrt{\frac{6}{\lambda}}\leq & a_o & <\infty.\\
\end{array}
\end{equation}
The equation for oscillating modes $a_k(\eta)$~(\ref{aoak}) can be reduced
to the equation for Bessel functions
\footnote
{Some properties of the Bessel functions are presented in
Appendix~\ref{ApB}~\cite{Abram}.}

\begin{equation}
\frac{d^2\xi(\tau_k)}{d\tau^2_k}+\frac{1}{\tau_k}\frac{d\xi(\tau_k)}{d\tau_k}
+\left(1-\frac{\nu^2}{\tau^2_k}\right)\xi(\tau_k)=0
\end{equation}
with the index $\nu=5/2$ by change of variables

\begin{equation} \label{tk}
\tau_k=k(1-\eta),\;\;a_k(\tau_k)=\xi(\tau_k)\sqrt{\tau_k}.
\end{equation}
The general solution for $a_k(\eta)$ is a linear combination

\begin{equation}  \label{apsi}
a_k(\eta)=\frac{1}{\sqrt{2}}
\left[A_k\psi_k(\eta)+A^\dagger_k\psi^*_k(\eta)\right],
\end{equation}
where $A_k,A^\dagger_k$ are coefficients, and $\psi_k(\eta),\psi^*_k(\eta)$
are expressed in terms of Hankel functions

\begin{equation} \label{psi}
\psi_k(\eta)=\frac{1}{2}\sqrt{\pi\hbar(1-\eta)}
H^{(1)}_{\frac{5}{2}}\Bigl(k(1-\eta)\Bigr),\;\;
\psi^*_k(\eta)=\frac{1}{2}\sqrt{\pi\hbar(1-\eta)}
H^{(2)}_{\frac{5}{2}}\Bigl(k(1-\eta)\Bigr).
\end{equation}
Functions $\psi_k(\eta),\psi^*_k(\eta)$ satisfy the orthogonality condition

\begin{equation}    \label{orth}
\psi_k(\eta)\frac{\partial}{\partial \eta}\psi^*_k(\eta)-
\psi^*_k(\eta)\frac{\partial}{\partial \eta}\psi_k(\eta)=i\hbar.
\end{equation}
Using this relation, the coefficients $A_k,A^\dagger_k$ can be written as
follows:

\begin{equation} \label{AA}
A_k=\frac{\sqrt{2}}{i\hbar}
\left(a_k(\eta)\frac{\partial}{\partial \eta}\psi^*_k(\eta)-
\psi^*_k(\eta)\frac{\partial}{\partial \eta}a_k(\eta)\right),\;\;\;
A^\dagger_k=-\frac{\sqrt{2}}{i\hbar}
\left(a_k(\eta)\frac{\partial}{\partial \eta}
\psi_k(\eta)-\psi_k(\eta)\frac{\partial}{\partial \eta}a_k(\eta)\right).
\end{equation}
The equality

\begin{equation}
A^*_k=A^\dagger_k
\end{equation}
holds as the field $a(\eta,x)$ has a real value.

\subsection{\it Quantization of oscillatory modes}

As zero mode $a_o(\eta)$ is a classical quantity in the accepted
approximation~(\ref{aoggak}), only oscillating modes $a_k(\eta)$ must be
quantized as independent variables. Due to the negative kinetic energy,
the conjugated momentum has an opposite sign with respect to time derivative
of the $a_k(\eta)$

\begin{equation}
\pi_k(\eta)=\frac{\partial L}{\partial \dot a_k(\eta)}=-2\dot a_k(\eta).
\end{equation}
The momentum operator is as follows (see Appendix~\ref{ApA})

\begin{equation}
\pi_k\;\to\;i\hbar\frac{\delta}{\delta a_k}.
\end{equation}
The commutator also differs from that in a usual case by sign

\begin{equation}
[a_k,\pi_{k'}]=-i\hbar\delta_{kk'}.
\end{equation}
The wave functional of the system is a product of the mode functionals
in the case of the absence of interactions

\begin{equation}
\Psi=\prod_{k}^{}\Psi_k(\eta),\;\;\;\;\Psi_k(t)\equiv\Psi_k[a_k,\eta].
\end{equation}
The Schr\"odinger equation for $\Psi_k(\eta)$

\begin{equation}  \label{Shr}
-i\hbar\frac{\partial}{\partial \eta}\Psi_k(\eta)
=H_k\Psi_k(\eta)
\end{equation}
with the Hamiltonian
\begin{equation} \label{qham}
H_k=-\frac{\pi^2_k}{4}-(k^2-\lambda a^2_o(\eta))a^2_k,
\end{equation}
where $a_o(\eta)$ is a known function~(\ref{aot}), can be solved by usual
ansatz~\cite{GuthPi}

\begin{equation} \label{funk}
\Psi_k(\eta)=C_k(\eta)\exp\left\{-2\Omega_k(\eta)a^2_k\right\},
\end{equation}
where a covariance $\Omega_k(\eta)$ is introduced

\begin{equation}
\Omega_k(\eta)=-\frac{i}{2\hbar}
\frac{\dot\varphi^*_k(\eta)}{\varphi^*_k(\eta)}.
\end{equation}
The equations for $C_k(\eta)$ and $\varphi^*_k(\eta)$ follow from the
Schr\"odinger equation~(\ref{Shr})

\begin{equation}  \label{sist}
\left\{\begin{array}{l}
\frac{\dot C_k(\eta)}{C_k(\eta)}=-\frac{1}{2}
\frac{\dot\varphi^*_k(\eta)}{\varphi^*_k(\eta)},\\
\\
\ddot \varphi^*_k(\eta)+(k^2-\frac{6}{(1-\eta)^2})\varphi^*_k(\eta)=0.
\end{array}\right.
\end{equation}
The solution for $\varphi_k(\eta)$ is analogous to this for
$\psi_k(\eta)$~(\ref{psi})

\begin{equation} \label{vphi}
\varphi_k(\eta)=\frac{1}{2}\sqrt{\pi\hbar(1-\eta)}
\left(c_1H^{(1)}_{\frac{5}{2}}\Bigl(k(1-\eta)\Bigr)
+c_2H^{(2)}_{\frac{5}{2}}\Bigl(k(1-\eta)\Bigr)\right)
\end{equation}
with the same normalization~(\ref{orth}) suggesting a condition

\begin{equation}  \label{cc}
|c_1|^2-|c_2|^2=1.
\end{equation}
The coefficient $C_k(\eta)$ in the wave functional~(\ref{funk})
can be expressed via function $\varphi_k(\eta)$

\begin{equation}
\Bigl|C_k(\eta)\Bigr|=\left(\pi|\varphi_k|^2\right)^{-1/4}
\end{equation}
according to~(\ref{sist}) and to normalization condition

\begin{equation}
\int\limits_{}^{}{\cal D}a_k\Psi^*_k(t)\Psi_k(t)=1.
\end{equation}
The real and imaginary parts of the covariance $\Omega_k(\eta)$ are as
follows
\footnote{
It must be stressed that the real part would be negative in case of usual
Schr\"odinger equation instead of~(\ref{Shr}) and, hence,
the solution~(\ref{funk}) would be non-normalazible.}

\begin{equation} \label{reim}
Re\,\Omega_k(\eta)=\frac{1}{4|\varphi_k|^2},\;\;\;
Im\,\Omega_k(\eta)=-\frac{1}{4\hbar|\varphi_k|^2}
\frac{\partial}{\partial \eta}|\varphi_k|^2.
\end{equation}
Hence, the functional $\Psi_k(\eta)$~(\ref{funk}) is determined. It corresponds
to the Bunch -- Davies vacuum~\cite{Bunch} of the scalar field in
the de Sitter space which is a
unique, completely de Sitter invariant vacuum state~\cite{GuthPi}.
As it was stressed in~\cite{FHJ}
\footnote{
The author thanks Professor R. Jackiw for this reference.},
owing to conformal invariance, the
situation is as in flat space-time, where a unique Poincar\'e invariant
vacuum exists.

We demand $A_k$~(\ref{AA}) to be an annihilation operator

\begin{equation}
A_k\Psi_k(\eta)=0.
\end{equation}
This equality is satisfied by choosing $c_1\!=\!1$ in solution
$\varphi_k(\eta)$~(\ref{vphi}), hence we have
$\varphi_k(\eta)\equiv\psi_k(\eta)$
($c_2=0$ from~(\ref{cc})).
The annihilation $A_k$ and creation $A^\dagger_k$ operators~(\ref{AA})
are the following:

\begin{equation}  \label{crdestr}
A_k=\frac{\psi^*_k(\eta)}{\sqrt{2}}
\left(4\Omega_k(\eta)a_k-\frac{i}{\hbar}\pi_k\right),\;\;\;
A^\dagger_k=\frac{\psi_k(\eta)}{\sqrt{2}}\left(4\Omega^*_k(\eta)a_k
+\frac{i}{\hbar}\pi_k\right).
\end{equation}
They satisfy the usual commutation relation

\begin{equation}
[A_k,A^\dagger_{k'}]=\delta_{kk'}
\end{equation}
which preserves a positive definitized norm of the state functional.

Using expressions for operators $a_k,\pi_k$ via operators
$A_k,A^\dagger_k$~(\ref{crdestr})

\begin{equation}
a_k=\frac{1}{\sqrt{2}}(\psi_k(\eta)A_k+\psi^*_k(t)A^\dagger_{k'}),
\end{equation}

\begin{equation}
\pi_k=i\hbar 2\sqrt{2}(\psi_k(\eta)\Omega^*_k(\eta)A_k-
\psi^*_k(\eta)\Omega_k(t)A^\dagger_k)
=-\sqrt{2}(\dot\psi_k(\eta)A_k+\dot\psi^*_k(\eta)A^\dagger_k),
\end{equation}
we can rewrite the Hamiltonian~(\ref{qham}) in terms of creation
and annihilation operators

\begin{eqnarray}
H&=&-\frac{1}{2}\left\{
A_kA_k\left(\dot\psi^2_k(\eta)+(k^2-\frac{6}{(1-t)^2})\psi^2_k(\eta)\right)
\right.\\
&&+(A^\dagger_kA_k+A_kA^\dagger_k)\left(
\dot\psi_k(\eta)\dot\psi^*_k(\eta)+(k^2-\frac{6}{(1-\eta)^2})
\psi_k(\eta)\psi^*_k(\eta)\right)\nonumber\\
&&\left. + A^\dagger_kA^\dagger_k\left((\dot\psi^*_k(\eta))^2
+(k^2-\frac{6}{(1-\eta)^2})(\psi^*_k(\eta))^2\right)
\right\}.
\nonumber
\end{eqnarray}
As functions $\psi_k(\eta)$ tend to oscillator functions when $t\to 0$,
this Hamiltonian tends to the oscillator Hamiltonian for large $k$.

It is interesting to find a time dependence for the vacuum fluctuations
and for the one-particle excitation energy. Fluctuations are determined
by real part of the covariance $\Omega_k(\eta)$~(\ref{reim})

\begin{equation}  \label{fluct}
\langle a^2_k \rangle =
\int\limits_{}^{}{\cal D}a_k\Psi^*_k(\eta)a^2_k\Psi_k(\eta)
=\frac{1}{2}|\varphi_k|^2=\frac{1}{8Re\,\Omega_k(\eta)}.
\end{equation}
The one-particle state

\begin{equation}
\Psi_{1k}(\eta)=A^\dagger_k\Psi_k(\eta)=
\sqrt{2}\frac{\psi_k(\eta)}{|\psi_k|^2}a_k\Psi_k(\eta),\;\;\;
\langle \Psi^*_{1k}(\eta)|\Psi_{1k}(\eta) \rangle = 1
\end{equation}
has the energy

\begin{equation}
E_{1k}=\langle \Psi^*_{1k}(\eta)|H_k|\Psi_{1k}(\eta) \rangle -
\langle \Psi^*_{k}(\eta)|H_k|\Psi_{k}(\eta) \rangle
= -\dot\psi_k(\eta)\dot\psi^*_k(\eta)
-\psi_k(\eta)\psi^*_k(\eta)\left(k^2-\frac{6}{(1-\eta)^2}\right),
\end{equation}
where the vacuum energy is subtracted. The potential part of the energy
shows that the oscillations of the mode with the wave number
$k\!>\!\sqrt{6}$ break down at the moment $\eta=T_k$
\footnote {
This coincides with the equation of motion for
$a_k(\eta)$~(\ref{aoak}) if the solution for the $a_o(\eta)$~(\ref{aot})
is taken into account. The presence of the non-oscillating modes with
($k\!<\!\sqrt{6}$) at the time $\eta=0$ is a sequence of the initial value
of the zero mode $a_o(\eta=0)\!=\!\sqrt{6/\lambda}\;$.
}

\begin{equation} \label{Time}
T_k=1-\frac{\sqrt{6}}{k}.
\end{equation}
At this moment the potential becomes non-oscillatory, the amplitude and
fluctuations of the mode begin to grow. Explicit time dependence of the
energy is the following

\begin{equation} \label{e1}
E_{1k}=-\frac{\hbar k}{2}\zeta^{-6}(2\zeta^6-6\zeta^4-9\zeta^2-18),\;\;\;
\zeta=k(1-\eta),
\end{equation}
where an analytic expression for the spherical Bessel function is used
(see Appendix~\ref{ApB}).
For small $\eta\ll 1$ and large $k\gg\sqrt{6}$ (that means $\zeta\gg 1$)
the excitation energy is like a negative oscillator energy $E=-\hbar k$.
As the time grows ($\eta\!\to\! 1$), the lower bound for oscillating
modes wave number $k$ rises.
After time $\eta=T_k$~(\ref{Time}) a moment comes
when the excitation energy changes the sign. It can be seen from expression for
$E_{1k}$~(\ref{e1}) in the limit $\zeta\ll 1$ ($\eta\!\approx\! 1$).


\section{Relation of the conformal factor to an inflaton field and to
cosmological perturbations}
\label{Rel}


In this section two issues are discussed. The former is to show how
to present conformal factor as an inflaton field.
The latter shows how fluctuations of the conformal factor are included into the
theory of the cosmological perturbations.

As it follows from~(\ref{fluct}),~(\ref{vphi}), vacuum fluctuations
of the field $a(x)$ are similar to those of the scalar field in the
de Sitter metric~\cite{GuthPi}. To show an equivalence of
their dynamical properties explicitly,
let us rewrite the equation for $a_k$~(\ref{aoak})
in proper time~(\ref{tpr}), simultaneously preparing a conformal
transformation~\cite{Chern} of the mode

\begin{equation}
a_k=a_o\phi_k.
\end{equation}
The resulting equation is the same as one for a scalar field~\cite{GuthPi}

\begin{equation}
\frac{d^2\phi_k}{dt^2_{pr}}+3H
\frac{d\phi_k}{dt_{pr}}+H^2\left(k^2e^{-2Ht_{pr}}-4\right)\phi_k=0,
\end{equation}
where $H$ is the Hubble parameter following from solution for $a_o$~(\ref{aot})

\begin{equation}
H=\frac{1}{a_o}\frac{da_o}{dt_{pr}}=\sqrt{\frac{\lambda}{12}}.
\end{equation}
Induced by zero mode mass term $\mu^2=4H^2$ fixes the Bessel functions
index $\nu=5/2$ in according with the result~\cite{GuthPi}
(formula (3.18)). Moreover, the solution for the zero mode of the scalar field
(formula (7.6) in~\cite{GuthPi}) coincides with the solution for
$a_o$~(\ref{aot}).

Let us show that small non-homogeneities of the conformal factor correspond to
those of the two independent gauge invariant functions, which represent
cosmological scalar metric perturbations~\cite{Mukh}. For spatially flat
Robertson -- Walker metric, scalar perturbations are represented by two
3-scalar functions $\Phi(\eta,x)$ and $\Psi(\eta,x)$ ($\Phi,\Psi\ll 1$)

\begin{equation}
ds^2=a^2(\eta)\left[(1+2\Phi(\eta,x))d\eta^2
-(1-2\Psi(\eta,x))\delta_{\alpha\beta}dx^\alpha dx^\beta\right].
\end{equation}
In terms of new variables

\begin{equation}
C(\eta,x)=\frac{1}{2}\left(\Phi(\eta,x)-\Psi(\eta,x)\right),\;\;\;\;
N(\eta,x)=\frac{1}{2}\left(\Phi(\eta,x)+\Psi(\eta,x)\right)
\end{equation}
the perturbed interval looks like

\begin{equation} \label{dspert}
ds^2=a^2(\eta)(1+2C(\eta,x))\left[(1+2N(\eta,x))d\eta^2
-(1-2N(\eta,x))\delta_{\alpha\beta}dx^\alpha dx^\beta\right].
\end{equation}
Function $C(\eta,x)$ represents conformal factor fluctuations, and
function $N(\eta,x)$ is a generalization of the Newtonian potential.
The trace of the equations for the cosmological perturbations without
matter is as follows ($N(\eta,x)\equiv 0$)

\begin{equation} \label{Ceqn}
\ddot C+2{\cal H}\dot C -\nabla^2C-2({\cal H}^2+\dot{\cal H})C=0
\end{equation}
where $\dot C\equiv {dC}/{d\eta},\;{\cal H}=\frac{1}{a}\frac{da}{d\eta}$,
$\nabla^2$ is Laplacian in the space metric $\delta_{\alpha\beta}$.
A relation between function $C(\eta,x)$ and conformal factor $a(\eta,x)$
follows from comparison of the form of the intervals~(\ref{dspert})
and~(\ref{dscf}). The latter is following in the present paper

\begin{equation}
ds^2=a^2(\eta,x)[d\eta^2-\delta_{\alpha\beta}dx^\alpha dx^\beta].
\end{equation}
Taking into account decomposition~(\ref{decompos}) of the conformal
factor $a(\eta,x)$, one can deduce

\begin{equation}
C(\eta,x)\approx\frac{1}{\sqrt{2}a_o(\eta)}
\sum_{\vec k}^{}[a^+_{\vec k}(\eta)\cos\vec k\cdot\vec x
+a^-_{\vec k}(\eta)\sin\vec k\cdot\vec x],\;\;\;\;
a(\eta)=\sqrt{2}a_o(\eta)
\end{equation}
where $a_o(\eta)$ is considered as a known function~(\ref{aot}).
Then eq.~(\ref{Ceqn}) for the function $C(\eta,x)$ completely
coincides with the eq.~(\ref{scc})  for modes $a_k(\eta)$
\begin{equation}
\ddot a_k+(k^2-\lambda a_o^2)a_k=0.
\end{equation}
This correspondence follows from the fact that the equation of motion for the
conformal factor is proportional to the trace of the Einstein equations
according to~(\ref{sled}).

So we have showed that the conformal factor (4-scalar)
represents a kind of the inflaton field and how it is connected
to gauge invariant scalar metric perturbations.


\section{Summary}


The main aim of this work is to get a consistent way to present
gravitation as a field living on the flat background and
to apply quantum field technique with the well known interpretation
(especially, a link between quantum and classical theories)
to cosmology.
It seems preferable to describe the expansion of the Universe
in the functional Schr\"odinger picture which is suitable for
non-stationary processes like a phase transition. Here the
problem is solved in the not self-consistent case.

The representation of the conformally flat space-time
with the cosmological constant as a self-interacting scalar field in
Minkowski world is based on the fact that only one essential metric variable
(a conformal factor) describes the investigated model due to the conformal
symmetry
\footnote
{There is an interesting relation: 15 (conformal group
Killing vectors) plus one (conformal factor) is equal to 16 (variables in an
arbitrary tensor $\phi_{\mu\nu}$ in four dimension space-time).}.
After extraction of the conformal factor from the metric, the
Einstein -- Hilbert
action looks like an action for the conformal scalar field
with a negative kinetic term and a selfi-nteraction inspired by a cosmological
constant.  The metric energy-momentum tensor for such a field is proportional
to the initial Einstein equations
\footnote
{Lorentz~\cite{Lorentz} and Levi-Civita~\cite{Levi}
proposed to regard the Einstein tensor $G_{ik}$
as a gravitational energy momentum tensor.
It seems that such approach may change the number of the degrees of freedom.
Instead of the Einstein
equations without second time derivatives
(constraints~\cite{DiracC}), we now have a quantum average of the
components of the conserved energy-momentum tensor.}.
Hence, a vacuum state (in which
energy-momentum tensor average vanishes) corresponds to classical space-time,
and the field average is a scale factor of the Universe. From other side,
mentioned condition provides a link to the Wheeler -- DeWitt approach.

A role of the negative energy in quantum theory is investigated here.
In this case the momentum and the velocity of the particle are contrarily
directed. If an action is considered as a wave phase, then the constant phase
wave will move against a momentum direction. Hence, a coefficient
between the momentum and the wave vector must be negative. Further,
the negative energy leads to the negative ``action"-variable in the
Hamilton -- Jacobi theory. And the quantum of this ``action" must be negative,
too. Meanwhile, the minus sign before the Planck constant does not
disturb the Dirac's prescription for the quantization of a classical system.
The changing in sign with respect to
usual case occurs, for example, in commutator, in momentum operator,
in quasi-classical wave function
\footnote
{The sign becomes the same as it was suggested by Linde~\cite{LindeJ}.}
and in the left hand side of the Schr\"odinger equation. Quantization
of the system with the negative energy includes standard steps
and a probability interpretation if one follows the procedure suggested
here (see appendix~\ref{ApA})
\footnote{
An indefinite metric does not appear here in contrast with the
Gupta -- Bleuler quantization~\cite{Schweber}. The latter is based on the
assumption that a state with the negative energy is not observable.
This is fixed by weaker Lorentz condition. Without it, one can not
exclude states with a negative norm, and hence the probability interpretation
is away.

When fields with positive and negative energy are included into a theory,
one has a difficulties with spontaneous creation of infinite number of
particles of both types in the finite space-time volume~\cite{HawkingEllis}.
This remark will not be directly applicable here, at least because
conformal factor describes physical space-time.

In any case, an investigation of possibility of negative gravitational
energy is necessary. It is well known that Hartle -- Hawking~\cite{Hart}
and Vilenkin tunneling~\cite{Vilen} wave functions suffers difficulties
following Wheeler -- DeWitt approach.}.

In frames of the suggested model, the classical Universe arises at a
stage of the non-stable quantum system evolution when zero mode of
the conformal factor becomes much greater then oscillating modes.
So the classical space is the vacuum state of this system.
The large field vacuum average (zero mode) plays the role of the metric
scale factor. Integral of motion for zero mode is the energy vacuum
average and must be put to zero to coincide with the corresponding Einstein
equation. In result we derive the correct description of the classical
Universe. The existence of the field vacuum average is a
sequence of the instability of the system
(the kinetic energy is negative and self-interaction energy depending of the
cosmological constant is positive).

The ``phantom" field Lagrangian~\cite{Caldwell,Chiba} has negative
kinetic and positive potential parts, too
\footnote
{Narlikar and Padmanabhan~\cite{Inflat}
pointed out the similarity between inflation and
the steady state model and gave references where negative energy
momentum tensor was proposed to drive the expansion and matter creation.}.
In contrast with the conformal
factor, this field is introduced only for getting a negative pressure
to explain the accelerating Universe expansion.

The mentioned above peculiarities distinguish the suggested model from
Thirring model~\cite{Thirring} describing gravity by a scalar field in
the field theoretical formulation of the General Relativity in a flat
space
\footnote{
Gravitation theories as field theories in Minkowski space are based on the
bimetrical formalism introduced by Rosen~\cite{Rosen} and
Papapetrou~\cite{Papa}. An arbitrary second rank tensor field $\phi^{\mu\nu}$
can be decomposed into direct sum of the irreducible representations of the
non-homogeneous Lorentz group of the spin 2,1 and 0~\cite{Fron,Barnes}.}.
In this kind of theories the Universe expansion can not be described because
the positive definite Hamiltonian is postulated and small deviations from
Minkowski space are supposed~\cite{Thirring}.

In the model introduced here, the quantum fluctuations are similar to those
in the inflation theories~\cite{GuthPi} because the equations of motion for
oscillating modes are the same in both theories.
The index of Bessel functions in the solution for the conformal factor is
fixed, because the ``mass" parameter is connected with the cosmological
constant which simultaneously determines the Hubble parameter.

Further it seems interesting to do the following. In the self-consistent
case~(\ref{scc}), renormalization group running of the interaction constant
$\lambda$ can be studied. It must be shown at non-zero temperature that there
is a phase transition  causing the emergence of the vacuum field average due
to quantum fluctuations. Next, the quantum properties of the open, flat and
closed Universes can be compared using the fact that the background metric
curvature scalar is presented in the conformal field action~(\ref{csfact}). A
possible generalization for an arbitrary metric can be investigated. The most
intriguing problem is to incorporate a matter into the model.


\acknowledgments


The author is grateful to Prof. V.V.Papoyan and Dr. S.Gogilidze for
encouragement at the beginning of this work.  It is a pleasure to thank Dr.
V.Smirichinskii for helpful and critical discussions.  This work was supported
by the Russian Foundation of Basic Research, Grant No 01--01--00708.


\appendix


\renewcommand{\theequation}{\thesection.\arabic{equation}}

\setcounter{equation}{0}


\section{Systems with negative energy}
\label{ApA}


In this Appendix the quantization of a system with the negative energy
is discussed on the example of an inverted harmonic oscillator. It is shown
that the negative energy causes the minus in front of Planck constant. However
we do not need to change the known quantization algorithm and interpretation
of a resulting quantum theory.

First of all, let us discuss a difference between a classical motion of a
positive particle (with the positive kinetic energy) and of a negative
particle
(with the negative kinetic energy). We describe them by $L_+$ and $L_-$
Lagrangians respectively with the same linear potential

\begin{equation}
L_+=\frac{m\dot
x^2_+}{2}-kx_+,\;\;\; L_-=-\frac{m\dot x^2_-}{2}-kx_-,\;\;\; \dot
x_\pm=\frac{dx_\pm}{dt},
\end{equation}
where $m\!>\!0,\;\;k\!>\!0$. Their accelerations and velocities are
contradictory to each other

\begin{equation}
\ddot x_-=-\ddot x_+=\frac{k}{m},\;\;\;
\dot x_-=-\dot x_+=\frac{k}{m}t.
\end{equation}
The positive particle moves towards decreasing the potential and
the negative particle moves towards increasing the potential

\begin{equation}
x_-=-x_+=\frac{k}{m}\frac{t^2}{2}.
\end{equation}
From the energy expressions

\begin{equation}
E_+=\frac{m\dot x^2_+}{2}+kx_+,\;\;\;
E_-=-\frac{m\dot x^2_-}{2}+kx_-
\end{equation}
it follows that the classically allowed regions are above the potential barrier
for a positive particle and under the barrier for a negative particle,
respectively (Fig. 1).

\begin{figure}[t]
\centerline{\epsfig{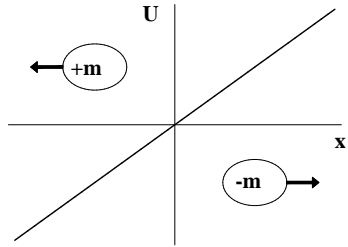}}
\caption{
The positive particle $+m$ moves towards decreasing the potential
above the potential barrier and
the negative particle $-m$ moves towards increasing the potential
under the barrier.
}
\label{fig1}
\end{figure}

The inverted oscillator has the following Lagrangian:

\begin{equation}
L=-\frac{m\dot x^2}{2}+\frac{m\omega^2x^2}{2}
\end{equation}
leading to the standard equation of motion

\begin{equation}   \label{urlagr}
\frac{d}{dt}\frac{\partial L}{\partial t}-\frac{\partial L}{\partial x}
=-m\ddot x-m\omega^2 x=0.
\end{equation}
With the momentum

\begin{equation} \label{imp}
p=\frac{\partial L}{\partial \dot x}=-m\dot x
\end{equation}
one derives the Hamiltonian (coinciding with energy)

\begin{equation} \label{ham}
H=p\dot x-L=-\frac{1}{2}\left(\frac{p^2}{m}+m\omega^2x^2\right).
\end{equation}
Hence, the energy is negative or zero

\begin{equation}
E=H\leq 0,
\end{equation}
and particle moves under the potential (Fig. 2).

\begin{figure}[t]
\centerline{\epsfig{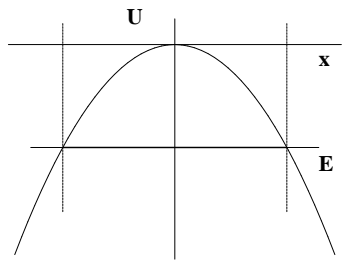}}
\caption{
A particle with negative energy $E$ oscillates under the potential.
}
\label{fig2}
\end{figure}

Using usual Poisson Bracket definition

\begin{equation} \label{CLSP}
\left\{A,B\right\}=
\frac{\partial A}{\partial x}\frac{\partial B}{\partial p}
-\frac{\partial A}{\partial p}\frac{\partial B}{\partial x},\;\;\;\;
\left\{x,p\right\}=1,
\end{equation}
for arbitrary function $A=A(x,p),\;B=B(x,p)$ one gets the Hamiltonian
equations

\begin{equation} \label{HAMEQ}
\left\{\begin{array}{l}
\dot x=\{x,H\}=-p/m,\\
\\
\dot p=\{p,H\}=m\omega^2x,
\end{array}\right.
\end{equation}
which correspond to the momentum definition~(\ref{imp}) and
to the Lagrangian equation~(\ref{urlagr}) but differ from the usual case by
its sign.

The straight corollary of the negative energy is the negative
``action"-variable in the Hamilton -- Jacoby theory where a momentum is
a partial derivative of the principal Hamilton function~\cite{Goldstein}

\begin{equation}
p=\frac{\partial S}{\partial x}.
\end{equation}
The Hamilton -- Jacoby equation

\begin{equation}  \label{GYA}
H\left(x,\frac{\partial S}{\partial x}\right)+\frac{\partial S}{\partial t}
=0
\end{equation}
with the negative oscillator Hamiltonian~(\ref{ham}) can be solved by
substitution

\begin{equation}
S(x,E,t)=W(x,E)-Et,
\end{equation}
where $W(x,E)$ is the characteristic Hamiltonian function. It follows from
this solution that the ``action"-variable is negative

\begin{equation} \label{J}
J=\oint pdx=\oint\frac{\partial W(x,E)}{\partial x}dx=\frac{2\pi}{\omega}E
\leq 0.
\end{equation}

So we have seen that two circumstances must be taken into
account at the quantization of a negative energy system.

First, an ``action"-quantum would be negative if it is considered as the
least possible portion of the ``action"~(\ref{J}) in the Bohr -- Zommerfeld
quantization. Let us suggest the following:

\begin{equation}
J=-nh,
\end{equation}
where a quantum number $n$ is positive or zero,
$h$ is the Planck constant.

Second, a surface~\cite{Goldstein}

\begin{equation}
S(\vec x,E,t)=W(\vec x,E)-Et=W_o=const
\end{equation}
moves along momentum $\vec p=\nabla W(\vec x,E)$ if $E\!>\!0$

\begin{equation}
S(t)=W_o=W_+-Et,\;\;\;\;W_o<W_+,
\end{equation}
and against it if $E\!<\!0$ (Fig. 3)

\begin{equation}
S(t)=W_o=W_--Et,\;\;\;\;W_o>W_- .
\end{equation}

\begin{figure}[t]
\centerline{\epsfig{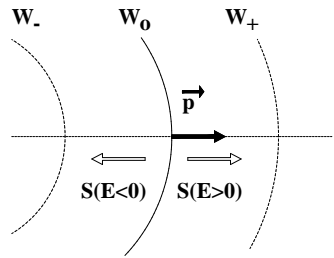}}
\caption{Figure.3
The wave $S(\vec x,E,t)=W_o$ moves along momentum $\vec p$
if $E\!>\!0$ and against it if $E\!<\!0$.
}
\label{fig3}
\end{figure}

Hence, the direction of a wave vector

\begin{equation}
\vec k=\nabla \Phi(\vec x,\omega,t),
\end{equation}
where $\Phi(\vec x,\omega,t)$ is a wave phase, coincides with the
direction of the momentum if $E\!>\!0$ and contradicts it if $E\!<\!0$

\begin{equation}
\left\{\begin{array}{l}
\vec k \uparrow \uparrow \vec p\;\;\;\;E>0,\\
\\
\vec k \uparrow \downarrow \vec p\;\;\;\;E<0.\\
\end{array}\right.
\end{equation}
Putting

\begin{equation}
\vec p=-\hbar \vec k,
\end{equation}
for the case $E\!<\!0$, one gets the following relation:

\begin{equation} \label{FAZA}
\Phi=-\frac{S}{\hbar}.
\end{equation}
Note that the wave vector coincides with the velocity as $\vec p=-m\vec v$,
and the circular frequency of the wave process is positive

\begin{equation}
\omega=\frac{\partial \Phi}{\partial t}=-\frac{1}{\hbar}
\frac{\partial S}{\partial t}=-\frac{E}{\hbar}>0,\;\;\;\;\mbox{where}\;E<0.
\end{equation}

Does the quantum mechanics allow a negative ``action"-quantum? According
to Dirac~\cite{Dirac} the main point in quantization of a classical system
is the replacement of a classical Poisson bracket by a quantum one. The latter
must be Hermitian as the former is real. The minus sign in front
of the Planck constant

\begin{equation} \label{CLQ}
\left\{A,B\right\}=\frac{1}{i(-\hbar)}(AB-BA)
=-\frac{1}{i\hbar}[A,B]
\end{equation}
does not disturb this condition. Using formulae~(\ref{CLSP}),~(\ref{CLQ}),
one gets a commutation relation

\begin{equation} \label{comut}
[x,p]=-i\hbar,
\end{equation}
where the momentum operator has the form

\begin{equation}
\hat p\equiv i\hbar\frac{\partial}{\partial x}.
\end{equation}
The quasi-classical wave function reads as following according to~(\ref{FAZA})

\begin{equation}
\Psi(x,t)=\exp\left\{-\frac{i}{\hbar}S(x,t)\right\}
\end{equation}
and must satisfy the Schr\"odinger equation

\begin{equation} \label{SHRED}
-i\hbar\frac{d}{dt}\Psi(x,t)=H(x,\hat p)\Psi(x,t).
\end{equation}
The energy operator changes the sign with respect to a usual case, too

\begin{equation}
E\to -i\hbar\frac{d}{dt}.
\end{equation}
The Schr\"odinger equation~(\ref{SHRED}) undergoes into the
Hamilton --Jacoby equation~(\ref{GYA}) in the classical limit $\hbar\to 0$.

The probability conservation law holds

\begin{equation}
\frac{d\rho}{dt}+\nabla\vec j=0,
\end{equation}
where

\begin{equation}
\rho=\Psi^*(x,t)\Psi(x,t),\;\;\;\;\;
\vec j=\frac{i\hbar}{2m}\left(\Psi(x,t)\nabla\Psi^*(x,t)-
\Psi^*(x,t)\nabla\Psi(x,t)\right).
\end{equation}
The quantity $\rho$ is positive and has a meaning of the probability
density under the normalizing condition

\begin{equation}
\int\rho dx=\int\Psi^*(x,t)\Psi(x,t)dx=1.
\end{equation}
The current $\vec j$ coincides with the wave vector

\begin{equation}
\vec j \uparrow \uparrow \vec k.
\end{equation}
The Heisenberg equations are the following:

\begin{equation}
\frac{d\hat A}{dt}=\frac{\partial\hat A}{\partial t}
-\frac{1}{i\hbar}[\hat A,H].
\end{equation}
Then, according to the definition of the average of the time derivative

\begin{equation}
\langle\frac{d\hat A}{dt}\rangle=\frac{d}{dt}\langle\hat A\rangle,
\end{equation}
one gets the Ehrenfest theorems

\begin{equation}
\left\{\begin{array}{l}
\frac{d}{dt}\langle\hat x\rangle
=\langle-\frac{1}{i\hbar}[\hat x,H]\rangle=-\langle \hat p\rangle /m,\\
\\
\frac{d}{dt}\langle\hat p\rangle
=\langle-\frac{1}{i\hbar}[\hat p,H]\rangle=m\omega^2\langle x\rangle,
\end{array}\right.
\end{equation}
which correspond to classical Hamiltonian equations~(\ref{HAMEQ}).

Now the Schr\"odinger picture and a filling number representation for
a negative harmonic oscillator can be considered. The Schr\"odinger
equation

\begin{equation}
(H-E)\Psi(x,t)
=\left(-\frac{1}{2}\left(\frac{\hat p^2}{m}+m\omega^2 x^2\right)-E\right)
\Psi(x,t)=0,
\end{equation}
or in coordinate representation

\begin{equation}
\left(\frac{\hbar^2}{2m}\frac{\partial^2}{\partial x^2}
-\frac{1}{2}m\omega^2x^2-E\right)\Psi(x,t)=0,
\end{equation}
differs in sign from the usual case but here the energy $E$ is less than
zero. Introducing dimensionless parameters

\begin{equation}
\xi=\frac{x}{x_d},\;\;\;x_d=\sqrt{\frac{\hbar}{m\omega}},\;\;\;
\varepsilon=\frac{-2E}{\hbar\omega}>0
\end{equation}
we get a usual equation

\begin{equation}
\left(\frac{\partial^2}{\partial \xi^2}-\xi^2+\varepsilon\right)\Psi(x,t)=0.
\end{equation}
Hence, the stationary states are characterized by negative energy levels

\begin{equation}
E_n=-\hbar\omega\left(n+\frac{1}{2}\right)
\end{equation}
and by wave functions

\begin{equation}
\Psi_n(x,t)=\frac{1}{\left(\sqrt{\pi}n!2^nx_d\right)^{1/2}}
\exp\left\{-\frac{1}{2}\left(\frac{x}{x_d}\right)^2\right\}
H_n\left(\frac{x}{x_d}\right),
\end{equation}
where Hermit polynomials

\begin{equation}
H_n(\xi)=(-)^ne^{\xi^2}\frac{d^n}{d\xi^n}e^{-\xi^2}
\end{equation}
satisfy the orthogonality condition

\begin{equation}
\int\limits_{-\infty}^{+\infty}\Psi^*_m(x,t)\Psi_n(x,t)dx=\delta_{mn}.
\end{equation}
To undergo to the filling number representation, let us introduce operators

\begin{equation}
a=\frac{1}{\sqrt{2}}\left(\xi+\frac{\partial}{\partial \xi}\right)
=\frac{1}{\sqrt{2}}\left(\sqrt{\frac{m\omega}{\hbar}}x
-i\frac{\hat p}{\sqrt{\hbar m\omega}}\right),
\end{equation}

\begin{equation}
a^\dagger=\frac{1}{\sqrt{2}}\left(\xi-\frac{\partial}{\partial \xi}\right)
=\frac{1}{\sqrt{2}}\left(\sqrt{\frac{m\omega}{\hbar}}x
+i\frac{\hat p}{\sqrt{\hbar m\omega}}\right),
\end{equation}
satisfying the commutation relation

\begin{equation}
[a,a^\dagger]=1
\end{equation}
according to~(\ref{comut}). The operator $a$ annuls the vacuum state

\begin{equation}
|0\rangle=\Psi_o(x,t)=\frac{1}{\pi^{1/4}}e^{-\xi^2/2}
,\;\;\;\;\langle 0|0\rangle=1,
\end{equation}

\begin{equation}
a|0\rangle=0.
\end{equation}
The operator $a^\dagger$ acting on to the vacuum
gives exited states with the quantum number $n$

\begin{equation}
|n\rangle=\Psi_n(x,t)=\frac{1}{\sqrt{n!}}(a^\dagger)^n\Psi_o(x,t),\;\;\;
\langle n|n\rangle=1.
\end{equation}
The energy operator looks as follows:

\begin{equation}
H=-\hbar\omega\left(\hat n+\frac{1}{2}\right),
\end{equation}
where a particle number operator is introduced

\begin{equation}
\hat n=a^\dagger a,\;\;\;\;\hat n |n\rangle=n |n\rangle.
\end{equation}
To give the interpretation of the operators $a,a^\dagger$ let us find
the value of the operators $H$ and $\hat n$ in the stations
$a^\dagger |n\rangle$, $a |n\rangle$

\begin{equation}
Ha^\dagger |n\rangle=(E_n-\hbar\omega)a^\dagger |n\rangle,\;\;\;\;
\hat n a^\dagger |n\rangle=(n+1)a^\dagger |n\rangle,
\end{equation}

\begin{equation}
H a |n\rangle=(E_n+\hbar\omega)a |n\rangle,\;\;\;\;
\hat n a |n\rangle=(n-1)a |n\rangle.
\end{equation}
The operator $a^\dagger$ decreases the energy onto the quantity $\hbar\omega$
and increases the particle number onto a unit. This operator can be called
the creation operator for the particle with the negative energy
$-\hbar\omega$.
The operator $a$ increases the energy onto the quantity $\hbar\omega$
and decreases the particle number onto a unit. This operator can be called
the annihilation operator for the particle with the negative energy
$-\hbar\omega$.

So the quantum theory of the negative energy system needs a negative
``action"-quantum value by preserving the usual apparatus and probability
interpretation.

\setcounter{equation}{0}


\section{Bessel functions}
\label{ApB}

The first- and second-order Hankel functions~\cite{Abram} are linear
combinations of the Bessel functions

\begin{equation}
\begin{array}{l}
H^{(1)}_\nu(\zeta)=J_\nu(\zeta)+iY_\nu(\zeta)\\
\\
H^{(2)}_\nu(\zeta)=J_\nu(\zeta)-iY_\nu(\zeta)\\
\end{array}
\end{equation}
and satisfy the following orthogonality relation:

\begin{equation}
H^{(1)}_\nu(\zeta)\frac{d}{d\zeta}H^{(2)}_\nu(\zeta)
-H^{(2)}_\nu(\zeta)\frac{d}{d\zeta}H^{(1)}_\nu(\zeta)
=-\frac{4i}{\pi\zeta}.
\end{equation}
They are complex conjugated by real argument $\zeta\in\Re$

\begin{equation}
\left(H^{(1)}_\nu(\zeta)\right)^*=H^{(2)}_\nu(\zeta).
\end{equation}
The Bessel functions $J_{\frac{5}{2}}(\zeta)$ и $Y_{\frac{5}{2}}(\zeta)$
can be expressed via spherical Bessel functions~\cite{Abram}

\begin{equation}
\begin{array}{l}
j_2(\zeta)=\sqrt{\frac{\pi}{2\zeta}}J_{\frac{5}{2}}(\zeta)
=\left(\frac{3}{\zeta^3}-\frac{1}{\zeta}\right)\sin(\zeta)
-\frac{3}{\zeta^2}\cos(\zeta),\\
\\
y_2(\zeta)=\sqrt{\frac{\pi}{2\zeta}}Y_{\frac{5}{2}}(\zeta)
=\left(-\frac{3}{\zeta^3}+\frac{1}{\zeta}\right)\cos(\zeta)
-\frac{3}{\zeta^2}\sin(\zeta).
\end{array}
\end{equation}



\begin{thebibliography}{99}

\bibitem{Sing} J.L. Sing, {\it Relativity: The General Theory}
              (Norht-Holland Publ. Comp., Amsterdam 1960).


\bibitem{DeWitt} B.S. DeWitt. Phys. Rev. {\bf 160}, 1113 (1967).

\bibitem{Isham} W.F. Blyth, and C.J. Isham,
                Phys. Rev. D {\bf 11}, 768 (1975).

\bibitem{Narlik1} T. Padmanabhan and J.V. Narlikar,
                  Gen. Rel. Grav. {\bf 13}, 669 (1981).

\bibitem{Pad1} T. Padmanabhan, Phys. Rev. D {\bf 28}, 745 (1983).


\bibitem{NarPad} J.V. Narlikar and T. Padmanabhan,
                  Phys. Rep. {\bf 100}, 151 (1983).


\bibitem{Hawking}  S.W. Hawking, in {\it General Relativity. An
		 Einstein Centenary Survey}, edited by S.W. Hawking and
                 W. Israel (Cambridge Univ. Press, Cambridge, 1979), p. 746.

\bibitem{Gibbons} G.W. Gibbons, S.W. Hawking, and M.J. Perry,
		Nucl. Phys. {\bf B138}, 141 (1978).


\bibitem{Schleich} K. Schleich, Phys. Rev. D {\bf 36} 2342 (1987).

\bibitem{LindeJ} A.D. Linde,
	     Zh. \'Eksp. Teor. Fiz. {\bf 87}, 369 (1984)
             [Sov. Phys. JETP {\bf 60}, 211 (1984)].

\bibitem{Weinbergbook} S. Weinberg, {\it Gravitation and Cosmology:
             Principles and Applications of the General Theory of
             Relativity} (New York: Wiley, 1972).


\bibitem{Ryan}{M. Ryan, {\it Hamiltonian Cosmology},
               Lecture Notes in Physics N 13,
              (Springer-Verlag, Berlin, 1972).}


\bibitem{Pad15} T. Padmanabhan, Gen. Rel. Grav. {\bf 15}, 435 (1983).

\bibitem{Pad19} T. Padmanabhan, Gen. Rel. Grav. {\bf 19}, 927 (1987).

\bibitem{Perl} S. Perlmutter et al., Astrophys. J. {\bf 517}, 565 (1999).
	       A.G. Riess at al., The Astronom. J. {\bf 116}, 1009 (1998).

\bibitem{WeinbL} S. Weinberg, Rev. Mod. Phys. {\bf 61}, 1 (1989).

\bibitem{Kolb} E.W. Kolb, M.S. Turner, {\it The Early Universe},
	       Frontiers in Physics {\bf 69}
               (Addison Wesley Publ. Comp., 1990).

\bibitem{Gonzalez} P.F. Gonz\'alez-D\'{\i}az, Phys. Rev. D {\bf 62},
                 023513 (2000).

\bibitem{Lyth} A.R.Liddle and D.H. Lyth, Phys. Rep. {\bf 231}, 1 (1993).
	       D.H. Lyth and A. Riotto, Phys. Rep. {\bf 314}, 1 (1999).

\bibitem{Halliwell} J.J. Halliwell and S.W. Hawking,
	       Phys. Rev. D {\bf 31}, 1777 (1985).

\bibitem{Sachs} R.H. Sachs and A.M. Wolfe, Astroph. J. {\bf 147}, 73 (1967).

\bibitem{Boyan} D. Boyanovsky, H.J. de Vega, R. Holman, and J. Salgado,
	       Phys. Rev. D {\bf 59}, 125009 (1999).

\bibitem{Jackiw} O. {\' E}boli, R. Jackiw, and So-Young Pi,
		  Phys. Rev. D {\bf 37}, 3557 (1988).
		  \\ R. Jackiw, {\it Analysis on infinite-dimensional
		  manifolds -- Schr\"odinger representation for
		  quantized fields}, S{\'e}minare de
		  Math{\' e}matiques Sup{\'e}rieures, Montr{\'
		  e}al, Qu\'ebec, Canada, June 1988 and V Jorge Swieca Summer
		  School, S\~ao Paulo, Brazil, January 1989, CTP\#1720 March
		  1989.

\bibitem{Chern} N.A. Chernikov, and E.A. Tagirov,
         Ann. Inst. H. Poincar\'e {\bf 9}A, 109 (1968).

\bibitem{Landau} L.D. Landau, and E.M. Lifshitz,
              {\it Classical Theory of Fields}
              (Pergamon Press, Oxford, 1975).

\bibitem{Misner}  C.W. Misner, K. Thorne, and J.A. Wheeler,
                {\it Gravitation} (Freeman, San Francisco 1973).



\bibitem{Eizen} L.P. Eisenhart, {\it Riemannian Geometry}
          (Princeton University, Princeton 1926).

\bibitem{Lightman} A.P. Lightman, W.H. Press, R.H. Price,
                   S.A. Teukolsky, {\it Problem book in
                   Relativity and Gravitation} (Princeton Univ.
                   Press, Princeton, New Jersey 1975).

\bibitem{Keane} A.J. Keane, and R.K. Barrett,
                   Class. Quantum Grav. {\bf 17}, 201 (2000).



\bibitem{Dubr} B.A. Dubrovin, S.P. Novikov, and A.T. Fomenko,
              {\it Modern  Geometry  Methods and Applications,
                Part II, The Geometry and Topology of Manifold }
               (Springer Verlag , New York 1984).

\bibitem{Gurs}   F. G\"ursey, in {\it Relativity, Groups and Topology},
		edited by C. DeWitt and B. DeWitt (New York, London 1964),
                 p. 91.

\bibitem{GuthPi} A.H. Guth and So-Young Pi,
                 Phys. Rev. D {\bf 32}, 1899 (1985).

\bibitem{Bunch} T.S. Bunch and P.C.W. Davies,
		Proc. R. Soc. Lond. {\bf A 360}, 117 (1978).

\bibitem{FHJ}  R. Floreanini, C.T. Hill, and R. Jackiw,
	       Ann. Phys. (N.Y.) {\bf 175}, 345 (1987).

\bibitem{Abram}  {\it Handbook of Mathematical Functions}
                 edited by M Abramowitz and I.A. Stegun
		(Dover, New York 1965).

\bibitem{Mukh} V.F. Mukhanov, H.A. Feldman, and R.H. Branderberger,
               Phys. Rep. {\bf 215}, 203 (1992).

\bibitem{Lorentz} H.A. Lorentz, Amst. Versl., Bd {\bf 23}, 1073 (1915),
Bd {\bf 24}, 1389 and 1759 (1916), Bd {\bf 25}, 468 and 1380 (1916).

\bibitem{Levi} T. Levi-Civita, $ds^2$ ensteiniani in campi newtoniani. I-IX//
	       Rend. Acc. Linc. (5). Bd {\bf 26} (1917), Bd {\bf 27} (1918),
	       Bd {\bf 28} (1919).

\bibitem{DiracC} P.A.M. Dirac, Canad. J. Math. {\bf 2}, 129 (1950);
                                Canad. J. Math.  {\bf 3}, 1 (1951);
                                Proc. Roy. Soc.  {\bf A246}, 326 (1958).

\bibitem{Schweber} S.S. Schweber, {\it An Introduction to Relativistic
		  Quantum Field Theory} (Harper and Row, N.Y. 1961).


\bibitem{HawkingEllis} S.W. Hawking, G.F.R. Ellis,
		  {\it The Large Scale Structure of Space-Time}
		   (Cambridge, Cambridge University Press, 1973).

\bibitem{Hart} J.B. Hartle, S.W. Hawking, Phys. Rev. D {\bf 28},
               2960 (1983).

\bibitem{Vilen} A. Vilenkin, Phys. Rev. D {\bf 37}, 888 (1988).

\bibitem{Caldwell} R.R. Caldwell, astro-ph/9908168.

\bibitem{Chiba} T. Chiba, T. Okabe, and M. Yamaguchi,
		  Phys. Rev. D {\bf 62}, 023511 (2000).

\bibitem{Inflat} J.V. Narlikar, T. Padmanabhan, Annu. Rev. Astron.
		Astrophys. {\bf 29}, 325 (1991).

\bibitem{Thirring} W. Thirring, Ann. of Phys. (N.Y.) {\bf 16}, 96 (1961).

\bibitem{Rosen} N. Rosen, Phys. Rev. {\bf 57}, 147 and 150 (1940).

\bibitem{Papa} A. Papapetrou, Proc. R. Irish Acad. {\bf A52}, 11 (1948).

\bibitem{Fron} C. Fronsdal, Nuovo Chimento Suppl. {\bf 9}, 416 (1958).

\bibitem{Barnes} K.J. Barnes, J. Math. Phys. {\bf 6}, 788 (1965).


\bibitem{Goldstein} H. Goldstein, {\it Classical Mechanics}
                    (Addison-Wesley Press, INC. Cambridge. Mass. 1950).

\bibitem{Dirac}  P.A.M. Dirac, {\it The Principles of Quantum Mechanics}
                 (Oxford at the Clarendon Press 1958).

\end{thebibliography}
\end{document}